\documentclass[conference]{IEEEtran}
\IEEEoverridecommandlockouts
\usepackage{cite}
\usepackage{amsmath,amssymb,amsfonts}
\usepackage{algorithmic}
\usepackage{graphicx}
\usepackage{textcomp}
\usepackage{xcolor}
\usepackage{tabularx,booktabs}
\def\BibTeX{{\rm B\kern-.05em{\sc i\kern-.025em b}\kern-.08em
    T\kern-.1667em\lower.7ex\hbox{E}\kern-.125emX}}
\begin{document}

\title{Comparing Conventional Pitch Detection Algorithms with a Neural Network Approach\\

\author{\IEEEauthorblockN{Anja Kroon \\ ECSE 523 Speech Communications Final Project}
\IEEEauthorblockA{\textit{Dept. of Electrical and Computer Engineering} \\
\textit{McGill University}\\
Montreal, Quebec, Canada \\
anja.kroon@mail.mcgill.ca}
}}

\maketitle

\begin{abstract}
 Despite much research, traditional methods to pitch prediction are still not perfect. With the emergence of neural networks (NNs), researchers hope to create a NN based pitch predictor that outperforms traditional methods . Three pitch detection algorithms (PDAs), pYIN, YAAPT, and CREPE are compared in this paper. pYIN and YAAPT are conventional approaches considering time domain and frequency domain processing. CREPE utilizes a data trained deep convolutional neural network to estimate pitch. It involves 6 densely connected convolutional hidden layers and determines pitch probabilities for a given input signal. The performance of CREPE representing neural network pitch predictors is compared to more classical approaches represented by pYIN and YAAPT. The figure of merit (FOM) will include the amount of unvoiced-to-voiced errors, voiced-to-voiced errors, gross pitch errors, and fine pitch errors.
\end{abstract}

\section{Introduction}
Pitch is a characteristic of speech that is derived from the fundamental frequency. It exists in voiced speech only and is the rate at which the chords in the vocal tract vibrate. Speech can have voiced (periodic) and unvoiced (aperiodic) segments. Thus, during the unvoiced sections of speech, pitch is not present. Pitch detection algorithms (PDAs) seek to recognize when speech is voiced or unvoiced and detect the pitch during the voiced segments. PDAs can be used in speech applications to identify speakers, determine intonation, and distinguish tones all in real time. The applications extend to auditory aids for the deaf, automatic score transcription in music processing, language translation and vocoder systems. Although much research has been done in the field of pitch predictors, the algorithm has proven very difficult to perfect resulting in a continuous stream of research. \par
Pitch prediction has been proven difficult because sound is ultimately produced from the human body -- an imperfect source. The periodicity produced by a person are often not perfectly periodic as seen in Fig. 1. \cite{b1}
\begin{figure}
    \centerline{\includegraphics[width=0.30\textwidth]{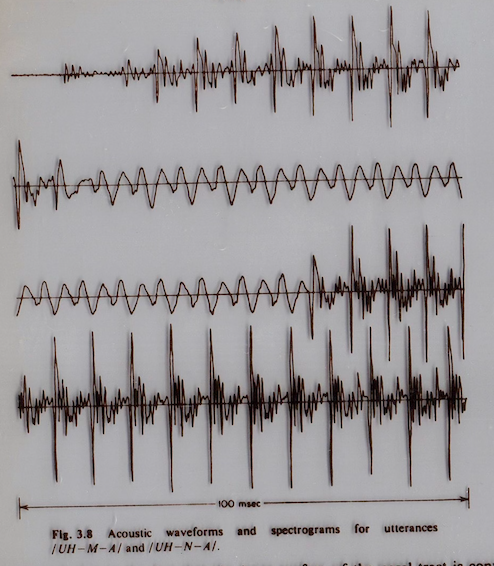}}
    \caption{Human Speech Waveform Displaying Imperfections in Periodicity and Transitions in Pitch Period \cite{b1}}
\end{figure}
This nature of the waveform makes it difficult to determine whether the signal is currently exhibiting periodic or aperiodic behavior. Secondly, because the sound is produced at the glottis, the waveform must travel through the vocal tract and propagate out of the lips. As the waveform travels through the vocal tract, the signal may pick up some disturbances in phase and spectral shape making it difficult to identify periodicity. Thirdly, distinguishing between unvoiced speech and very low frequency voiced speech has proven challenging. This effect is exacerbated when speech undergoes rapid transitions between unvoiced and low frequency voiced speech. Lastly, even with speech that has easily identifiable periodic segments, it can be difficult to distinguish exactly where one segment ends and another begins. Humans may be able to make judgements and identify segments with similar pitch periods, see Fig. 1, but for a computer, this task has proven difficult.  PDA researchers today are continuously working to fix these issues to achieve better applications.
The main types of pitch predictors are those that process in the time domain, frequency domain, or a blend of both. With the emergence of neural networks, a fourth class of PDAs now exists based on machine learning. This paper will present three monophonic pitch predictor algorithms: pYIN (a time domain based algorithm), YAAPT (a time domain and frequency domain based algorithm), and CREPE (a deep convolutional neural network based algorithm). In this paper, we will compare the three pitch detection algorithms based on their voicing decision errors, gross pitch errors (defined as errors in pitch period values exceeding 1 ms), fine pitch errors (errors in pitch period values less than 1 ms), and the mean and standard deviation of these fine pitch errors. \cite{b2} \par
Section 2 will delve into the specifics of the three pitch predictor algorithms being investigated. Section 3 will describe the test data the experiments are being conducted upon. In Sections 4 and 5, the methods of comparison and results and discussion will be presented, respectively. Section 6 will summarize the conclusions.
\section{Pitch Detection Algorithms}
The three PDAs presented in this section are pYIN, YAAPT, and CREPE. These algorithms form the basis of this experiment comparing classical pitch detection algorithms with a newer neural network approach.

\subsection{pYIN}
 pYIN (Probabilistic YIN) is an improved version of the well known PDA YIN \cite{b4}. pYIN is a robust and effective time-domain pitch estimator based on the auto-correlation method. It has the capability to find higher values pitches efficiently and is well suited for high pitched voices and music. Although pYIN is based on the auto-correlation method, the algorithm includes many refinements to make the estimate more robust. These refinements include a difference function, proper normalization, improved thresholding and parabolic interpolation to reduce the impact of lower sampling rates. For details, the reader is referred to \cite{b4}. \par

Since pYIN outputs only one pitch value per frame, as seen in Fig. 2, it limits the options for smoothing the output pitch contour. pYIN modifies the original algorithm of YIN by having it output multiple pitch candidate values with its associated probabilities. A comparison of YIN and PYIN is shown in Fig. 2. This allows for determining an output contour by using these combined probabilities to determine the optimal path. The approach used is based on a Hidden Markov Model. In \cite{b3} it is reported that pYIN is outperforming the conventional YIN algorithm, including reduction in pitch doublings and voicing detections.
\par 
\begin{figure}
    \centerline{\includegraphics[width=0.45\textwidth]{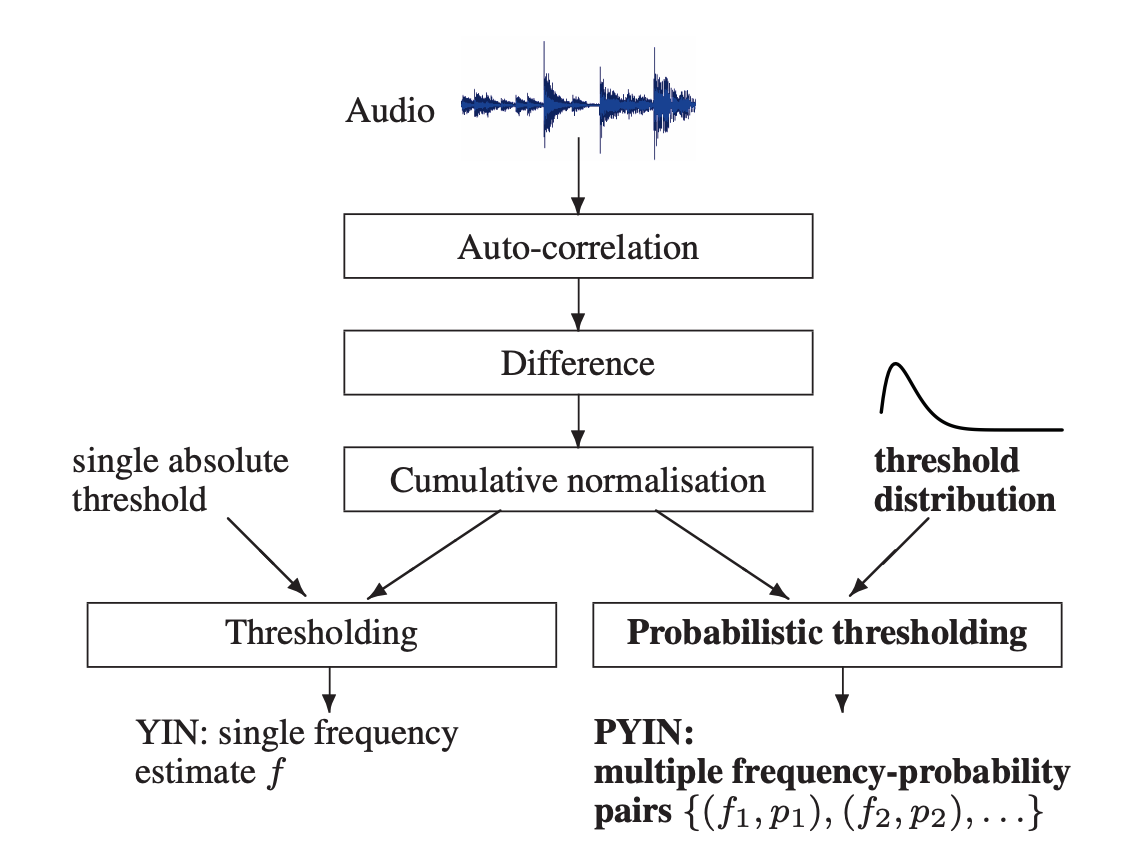}}
    \caption{Comparison of YIN and pYIN Algorithms}
\end{figure}

\subsection{YAAPT}
YAAPT stands for "yet another algorithm for pitch tracking". \cite{b5} The algorithm is based on a combination of time and frequency domain processing. It is based on an earlier algorithm called RAPT -- robust algorithm for pitch tracking \cite{b10}. RAPT employs the normalized cross correlation function (NCCF) which operates in the time domain.  Although RAPT is effective, it does incur a large amount of gross pitch errors due to frequent pitch doubling. To correct for this, YAAPT employs additional frequency domain analysis resulting in a more accurate pitch predictor. \par
YAAPT can be divided into 4 stages: preprocessing, pitch track calculation, pitch candidate estimation, and final pitch determination. Preprocessing involves making two signals that will then both be fed individually into the NCCF algorithm. The first is a bandpass filtered version of the original signal including frequencies between 50–1500 Hz. The second is a nonlinearly processed version of the original signal. Particularly, the absolute value and the  square of the signal were calculated. \par
Pitch track calculation is the next stage and includes analyzing the spectrogram of the two signals derived in preprocessing. The fundamental frequency, or pitch, is calculated directly from the spectrogram using spectral harmonics correlation and the normalized low frequency energy ratio. \par
The pitch is derived again using the NCCF based on the two signals derived in preprocessing forming the third stage, pitch candidate estimation. Cross referencing the pitch estimates from the spectrogram step, the algorithm is able to accurately determine the pitch and voicing of a desired signal using dynamic programming. A flow chart describing the steps of YAAPT in further detail is included in Fig. 3.
\begin{figure}
    \centerline{\includegraphics[width=0.45\textwidth]{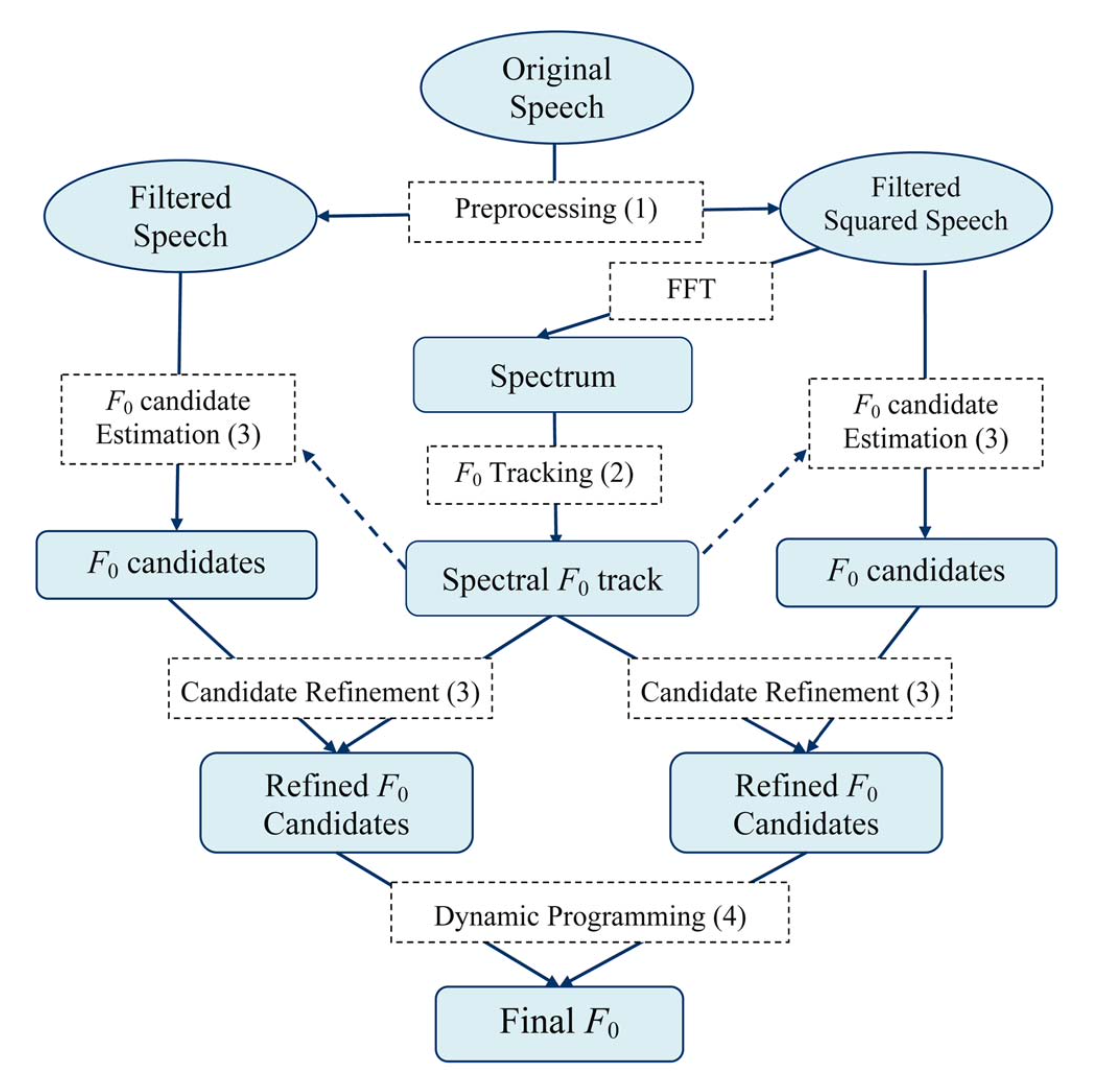}}
    \caption{Flow chart for YAAPT }
\end{figure}

\subsection{CREPE}

\begin{figure*}
    \centering
    \includegraphics[width=0.8\textwidth]{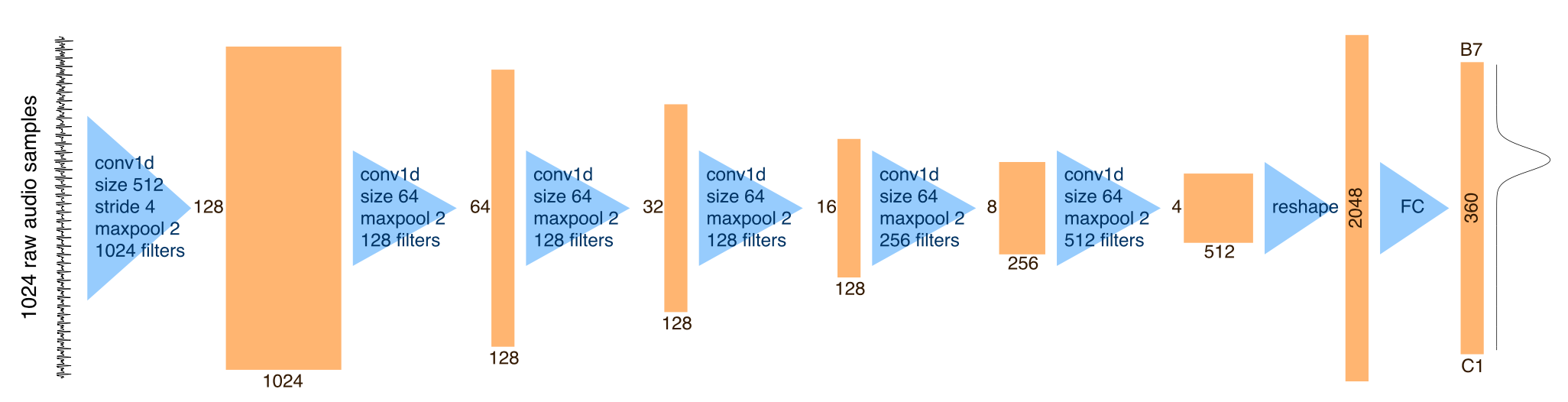}
    \caption{Flow Chart for CREPE.}
\end{figure*}

CREPE stands for ”convolutional representation for pitch estimation”. \cite{b6} It employs a deep convolutional neural network based algorithm and operates on the time domain waveform to determine pitch. The sampling rate is 16 kHz and the algorithm uses a 64 ms analysis window to process every 10 ms. The neural network in CREPE employs 6 densely connected convolutional layers. The neural network nature also indicates the pitch estimate will be determined via probabilities in the output layer. A flow chart describing the CREPE architecture is provided in Fig. 4. In contrast, to conventional PDAs such as YIN and YAAPT that analyze signal properties such as periodicity, spectral content, amplitude changes, CREPE does not do this analysis in a direct way.   Instead, the weights in the network are trained to produce for a given input an output that matches the provided correct answer from its training data base. This means that content and size of the training data base as well as the details of the training parameters will have a strong impact on its final performance. After training, the network is input input speech and it uses the weights obtained during training to produce an output. 
The data bases used for training of CREPE included exclusively synthetic music and no real voice samples. This likely will impact its performance for voice signals, but retraining the network for voice signals is outside the scope of this work.

\section{Test Data}
The aforementioned algorithms will be run on a pitch tracking database developed by Graz University of Technology. \cite{b7} The database includes microphone signals and reference pitch trajectories derived from larynx signals. In Fig. 5, an example microphone signal is shown in contrast with its laryngogram (measuring the larynx signal). There is clear periodicity shown in the laryngogram and pitch is thus easily identified. In contrast, the time signal has amplitude variation and formant effects that would make it much more difficult to estimate the periodicity. Further, when the laryngogram signal is once differentiated, the signal become a near perfect pulse train thus making it even easier to determine the periodicity. \cite{b8} \par
\begin{figure}
    \centerline{\includegraphics[width=0.45\textwidth]{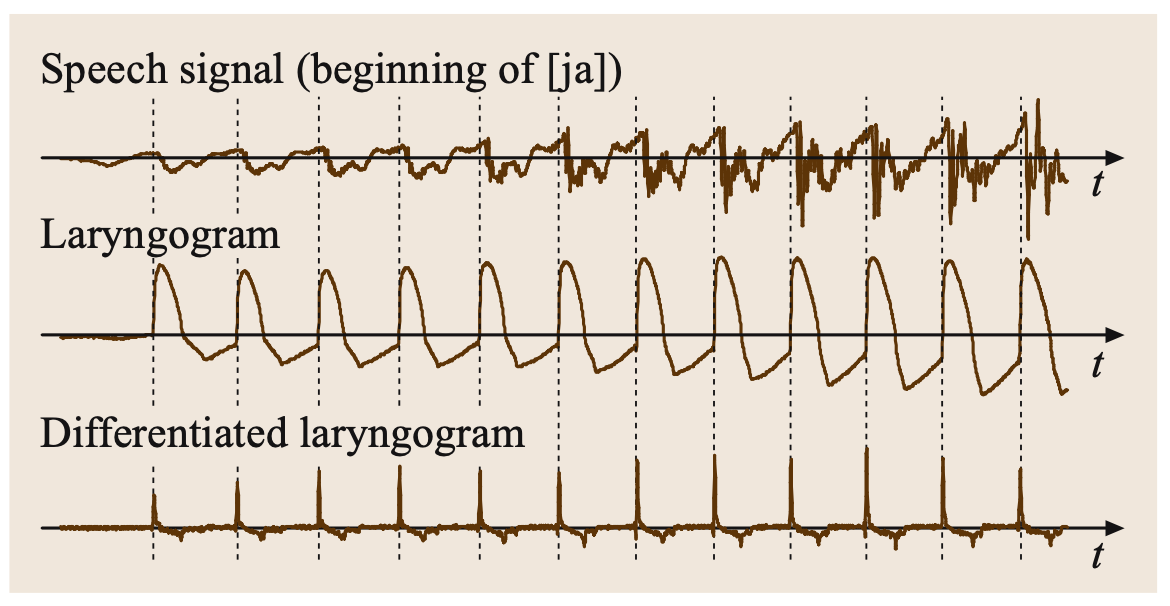}}
    \caption{Microphone Signal versus Laryngogram versus Differentiated Laryngogram \cite{b8}}
\end{figure}

The microphone signals in the database are digitized at 48 kHz and have 16 bit resolution. These reference pitch trajectories are thus considered to be the ground truth. The reference pitch trajectories are computed every 10 ms. The input microphone signals used include 10 sentences spoken by female speakers and 10 sentences spoken by male speakers. The speaker profiles are provided in Table 1. The speakers are diverse in age and accent. The 10 recorded utterances, their voiced frames, minimum, median, and maximum pitch values are provided in Table 2. The minimum pitch is 81.8 Hz and the maximum pitch is 328.3 Hz. All PDAs considered are able to determine pitch in this range. The utterances used in this experiment are all classified as phonetically diverse and diverse in phonetic context providing a robust test case for the PDAs. \par

\begin{table}[ht]
\caption{Speaker Profiles}
\begin{center}
\begin{tabular}{|c|c|c|c|}
\hline
\multicolumn{1}{|c|}{Speaker ID}
& \multicolumn{1}{|c|}{Age}
& \multicolumn{1}{|c|}{Sex}
& \multicolumn{1}{|c|}{Home Country}\\
F01      &  40  &  Female  &  Ireland   \\ \hline
F02      &  25  &  Female  &    USA  \\ \hline
F03      &  22  &  Female  &    Canada  \\ \hline
F04      &  26  &  Female  &    Canada \\ \hline
F05     &  48 & Female & USA \\ \hline
F06      & 28 & Female & USA \\ \hline
F07      & 24 & Female & USA \\ \hline
F08      & 22 & Female & England \\ \hline
F09      & 22 & Female & USA \\ \hline
F10      & 35 & Female & USA \\ \hline
M01      &  24  &  Male  &  South Africa   \\ \hline
M02      &  40  &  Male  &    England \\ \hline
M03      &  35  &  Male  &    England  \\ \hline
M04      &  26  &  Male  &    USA \\ \hline
M05     &  25 & Male & England \\ \hline
M06      & 23 & Male & USA \\ \hline
M07      & 24 & Male & USA \\ \hline
M08      & 24 & Male & England \\ \hline
M09      & 24 & Male & Canada \\ \hline
M10      & 33 & Male & USA\\ \hline
\end{tabular}
\label{table-tab2}
\end{center}
\end{table}

\begin{table*}
     \caption{Description of Utterances}
    \label{my-label}
    \begin{tabularx}{\textwidth}{@{}l*{10}{c}c@{}}
    \toprule
    Test & Test & Voiced& Min & Median & Max  \\ File & Sentences & Frames & Pitch [Hz] & Pitch [Hz] &Pitch [Hz] \\
    \midrule
    ./mic\_F01\_si453\_f0.txt   &    Everything went real smooth the sheriff said.   &  168   &  81.8 & 188.5 & 284.6  \\ 
    ./mic\_F02\_si643\_f0.txt & The armchair traveller preserves his illusions. & 163	&155.6&	191.8&	298.6 \\
    ./mic\_F03\_si831\_f0.txt&	Selecting banks by economic comparison is usually an individual problem.&	281&	168.9&	198.5&	232.4 \\ 
    ./mic\_F04\_si1020\_f0.txt	&The cowboy called this breed of cattle magpies.&	97	&76.6&	172.6	&206.5  \\ 
    ./mic\_F05\_si1209\_f0.txt	&If disobeyed, the result is turmoil and chaos.&	190	&98.3&	189.4&	229.0\\ 
    ./mic\_F06\_si1398\_f0.txt&	Do they make class biased decisions?&	153	&167.5	&212.1&	328.3 \\ 
    ./mic\_F07\_si1587\_f0.txt	&The jaded amorist conjured up pictures of blasphemous rights with relish. &	225&	65.8&	197.7&	231.7\\ 
    ./mic\_F08\_si1776\_f0.txt	&Afraid you'll lose your job if you don't keep your mouth shut?&	142	&105.0&	236.2&	326.5 \\
    ./mic\_F09\_si1965\_f0.txt&	If the other pilots were worried, they did not show it. &	177	& 172.6&	208.2&	310.4 \\
    ./mic\_F10\_si2154\_f0.txt	&She must have put his clothes in the closet.&	80&	157.5	&168.7&	209.1\\
    \addlinespace
    ./mic\_M01\_si453\_f0.txt	&Everything went real smooth the sheriff said.	&142	&94.7	&118.7&	181.7\\
    ./mic\_M02\_si642\_f0.txt	&Their props were two stepladders, a chair, and a palm fan.	&148&	54.4	&100.2	&153.5 \\
    ./mic\_M03\_si831\_f0.txt	&Selecting banks by economic comparison is usually an individual problem.&	268	&54.6&	85.9	&120.1\\
    ./mic\_M04\_si1020\_f0.txt	&The cowboy called this breed of cattle magpies.&	170&	87.8&	110.8&	152.1\\
    ./mic\_M05\_si1209\_f0.txt	&If disobeyed, the result is turmoil and chaos.	&160&	84.7&	130.4	&245.2\\
    ./mic\_M06\_si1398\_f0.txt &	Do they make class biased decisions?&	139&	104.9&	116.6	&155.5\\
    ./mic\_M07\_si1587\_f0.txt&	The jaded amorist conjured up pictures of blasphemous rights with relish. &	227&	105.5&	138.6&	190.2\\
    ./mic\_M08\_si1776\_f0.txt&	Afraid you'll lose your job if you don't keep your mouth shut?&	118&	85.0&	108.7&	127.5\\
    ./mic\_M09\_si1965\_f0.txt	&If the other pilots were worried, they did not show it. &	148&	58.8&	109.1&	185.4\\
    ./mic\_M10\_si2154\_f0.txt&She must have put his clothes in the closet.&	106&	84.3&	105.9&	150.7\\
    \bottomrule
    \end{tabularx}
    \end{table*}

Upon inspection of a microphone signal's waveform as seen in Fig. 6, a couple things must be noted. First, the waveform includes many seconds of silence. This silence was not removed from the test data as the unvoiced sections would always return a pitch of 0. Any other value for this unvoiced section would be considered an error. Additionally, the amplitude varies rapidly and over a large range -- another indication of a robust test signal. For this same waveform, the histogram is plotted in Fig. 7. The histogram of the reference pitch values provides a reference point for typical pitch values in a microphone signal for this specific test signal. The pitch includes some low frequency components below 100 Hz but mostly concentrates the pitch measurements between 100 and 300 Hz. This range is typical for a female speaker. With input data collected and representative of typical speech, comparison of PDA performance may begin.

\begin{figure}
    \center
    \includegraphics[width=0.45\textwidth]{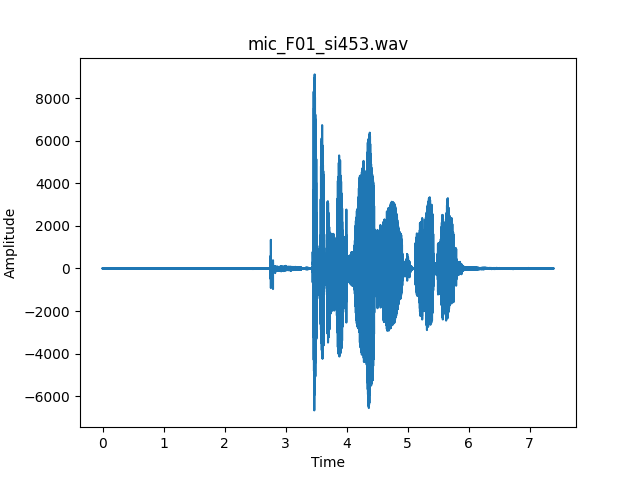}
    \caption{Example Waveform of a Microphone Signal}
\end{figure}

\begin{figure}
    \center
    \includegraphics[width=0.45\textwidth]{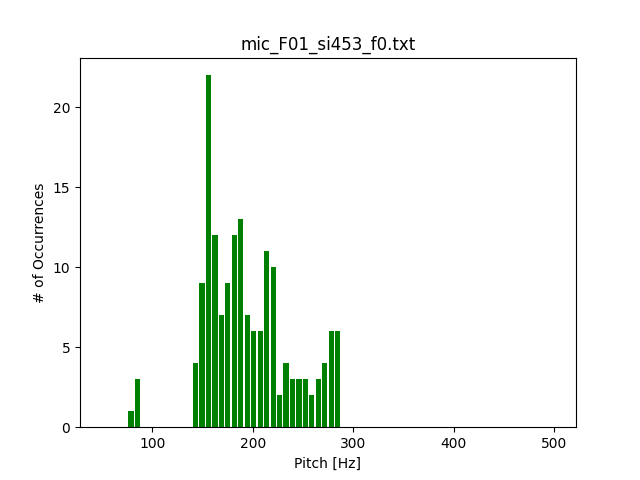}
    \caption{Histogram of Reference Pitch Values of the Microphone Signal of Figure 6}
\end{figure}

\section{Methods of Comparison}
The PDAs are compared based on their voicing decisions, gross pitch errors and fine pitch errors. These methods of comparison are derived from the Rabiner paper \cite{b2}.

\subsection{Voicing Decisions}
There are two types of voicing errors voiced-to-unvoiced and unvoiced-to-voiced. In the former, a voiced segment is misclassified as unvoiced. In the latter, an unvoiced segment is misclassified as voiced. \cite{b2} Unvoiced segments in the context of pitch are represented by the value 0.0 Hz. The importance of error type will depend on the application. For parametric voice coders, the pitch track shall be used to reconstruct the original microphone signal. A voiced-to-unvoiced misclassification would cause great disturbance in the output signal and would likely be perceptive to the listener. Weight of error is solely dependent on the type of applications. For this experiment, the voiced-to-unvoiced and unvoiced-to-voiced errors are weighted equally in the FOM.

\subsection{Gross vs. Fine Pitch Errors}
The inverse of the pitch is the period — or the time in which the signal takes to repeat itself. As defined in Rabiner \cite{b2}, gross pitch errors are classified in the sense of time. That means if the PDA calculates the pitch to be more than 1 ms different than the reference pitch trajectory, a gross pitch error has occurred. If the calculated pitch is less than 1 ms different compared to the reference pitch trajectory, a fine pitch error has occurred. \par
Often, the reason why gross pitch errors occur is pitch doubling. Pitch doubling is when the period is perceived to be two periods rather than one. Pitch doubling can also occur when the energy of the fundamental frequency is small in comparison to the first formant thus confusing the pitch predictor. These gross pitch errors are typically outliers in the pitch tracking data and are thus removed from the data set before determining other comparison parameters. Particularly, the mean and standard deviation of the error is only calculated considering the fine pitch errors. \cite{b9} \par
\subsection{Experimental Setup}
The three PDAs, pYIN, YAAPT, and CREPE, are run with the 10 female and male test utterances from the test data described in Section 3. Care was taken to run the 3 predictors with similar parameters, such as min f0, max f0, framesize and centering of the analysis frame. Since the CREPE outputs always a pitch value and a confidence level between 0 and 1, we used a confidence threshold of 0.5 as a voicing decision. The resulting pitch tracks are then compared against the reference pitch trajectories also provided by the database. For each test utterance the PDA is applied to, the total frames, number of unvoiced frames, number of voiced frame, number of voicing errors (unvoiced to voiced and voiced to unvoiced), gross pitch errors, and fine pitch errors are recorded. After all data has been acquired for a specific PDA, the total number of frames processed, total number of unvoiced frames, total number of voiced frames, total number of voicing errors, total number of gross pitch errors, mean of fine pitch errors, and the standard deviation of the fine pitch errors are recorded. These parameters are then compiled together and used to compare the performance of the various PDAs.

\section{Results \& Discussion}
The results of the PDA comparison can be divided into two parts: voicing decisions and pitch accuracy. For the voicing decisions, the results are shown in Fig 9. YAAPT makes the least amount of voicing errors. YIN an CREPE are similar in terms of voiced to unvoiced errors. pYIN has almost twice the amount of unvoiced to voiced errors compared to CREPE and YAAPT. It is also remarkable that the neural net based solution CREPE is able to work well on voice signals despite voice not being part of the training part. Recall, the neural net approach, CREPE, was trained on music. \par
Gross pitch errors are rarely observed and are present only in the CREPE algorithm. With 16k Hz input, CREPE is the algorithm that provides the limitation on pitch resolution. The mean and standard deviation of fine pitch errors are therefore normalized to equivalent samples taken at 16k Hz. Fine pitch errors are also relatively small with CREPE and YIN having the same level of accuracy. YAAPT slightly worse but still reasonable. 

To compute the FOM, we quantify the results in bins. For the voicing errors, we consider the unvoiced to voiced errors as a percentage of the total unvoiced frames. We also consider the opposite scenario -- voiced to unvoiced errors as a percentage of the total voiced frames. If the percentage is less than 8, rank 1 is assigned. If the percentage is greater than 8 but less than 16, rank 2 is assigned. If the percentage is greater than 16, the rank 3 is assigned in the FOM. \par

When factoring the fine pitch errors into the FOM, bins are also defined. Both mean and standard deviation are considered. For the mean fine pitch errors, errors occurring in less than 0.5 samples are given rank 1. Errors greater than 0.5 but less than 1 samples are given rank 2. Errors greater than 1 sample are given rank 3. For the standard deviation of the fine pitch errors, we consider errors less than 8 (0.5 ms) samples as rank 1, greater than 8 but less than 16 (1ms) samples as rank 2 and everything above 16 samples as rank 3. Because gross pitch errors are rarely present and only occur in CREPE, they are excluded from the FOM. \par

Based on the FOM, it can be seen that all pitch predictors are close in ranking with YAAPT outperforming pYIN by 1 point and pYIN outperforming CREPE by 1 point.

\begin{table*}
    \caption{Final Comparison Results}
    \begin{tabularx}{\textwidth}{@{}l*{10}{c}c@{}}
    \toprule
    PDA & Total & Unvoiced & Unvoiced/Voiced & Voiced & Voiced/Unvoiced & Gross Pitch & Fine Pitch & Mean Fine & Stdev Fine & FOM \\ Name & Frames & Frames & Errors as \% of & Frames & Errors as \% of &  Error & Error & Pitch Error & Pitch Error\\ & & & Unvoiced Frames &  & Voiced Frames\\
    \midrule
    CREPE&	14477&	10411&	7\% &	2783&	19\%&	3&	2780&	0.96&	7.44 & 7\\
    YAAPT & 14477&	10552&	6\% &	3177&	4\% &	0&	3177&	0.29&	12.42 &5\\
    pYIN&	14457&	9882&	13\%&	2890&	14\%&	0&	2890&	0.44&	6.18 &6 \\
    \bottomrule
    \end{tabularx}
    \end{table*}

\begin{figure}
    \center
    \includegraphics[width=0.45\textwidth]{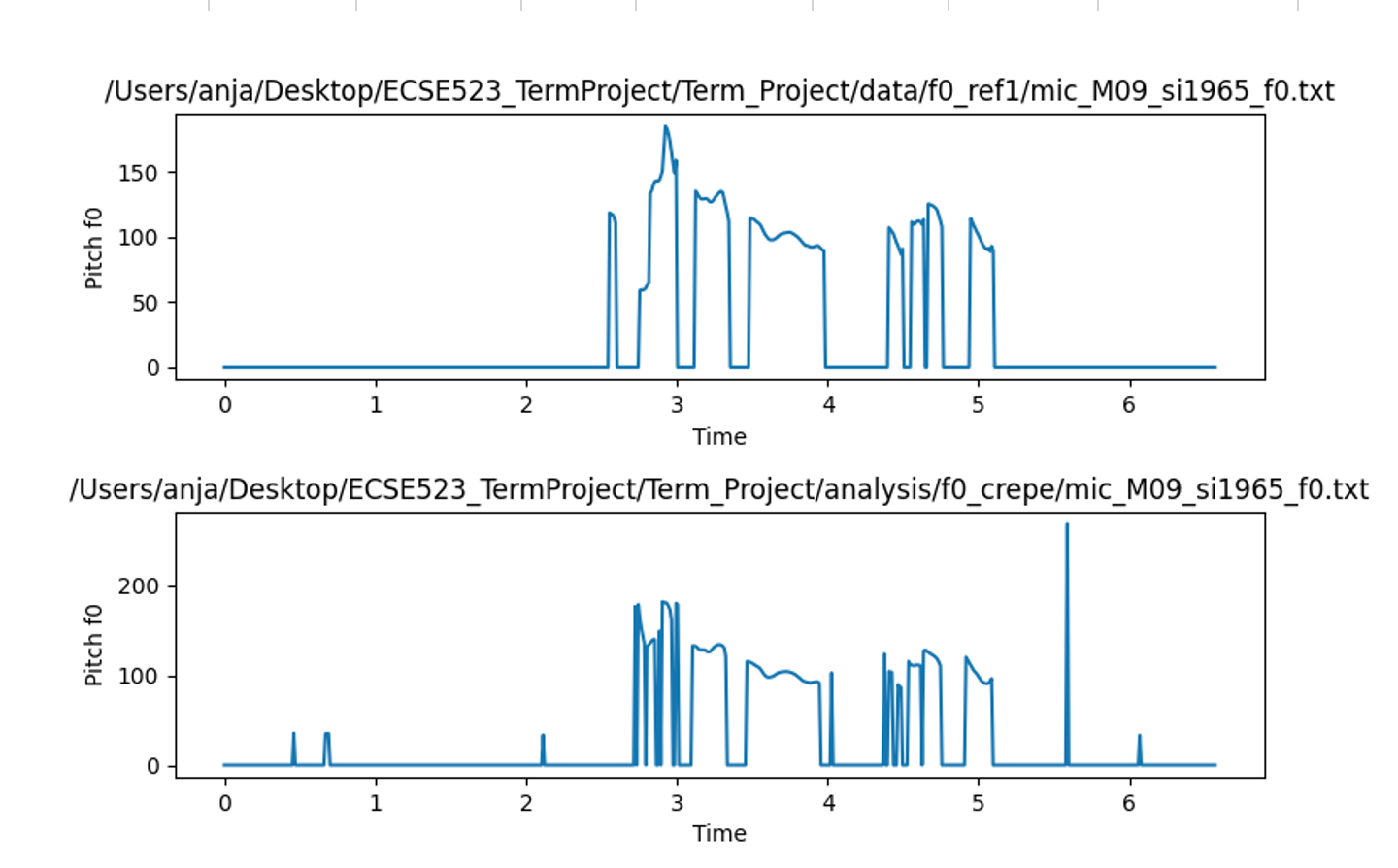}
    \caption{Example of Gross Pitch Errors Comparing Reference Pitch Track to CREPE Output Pitch Track }
\end{figure}
\begin{figure}
    \center
    \includegraphics[width=0.45\textwidth]{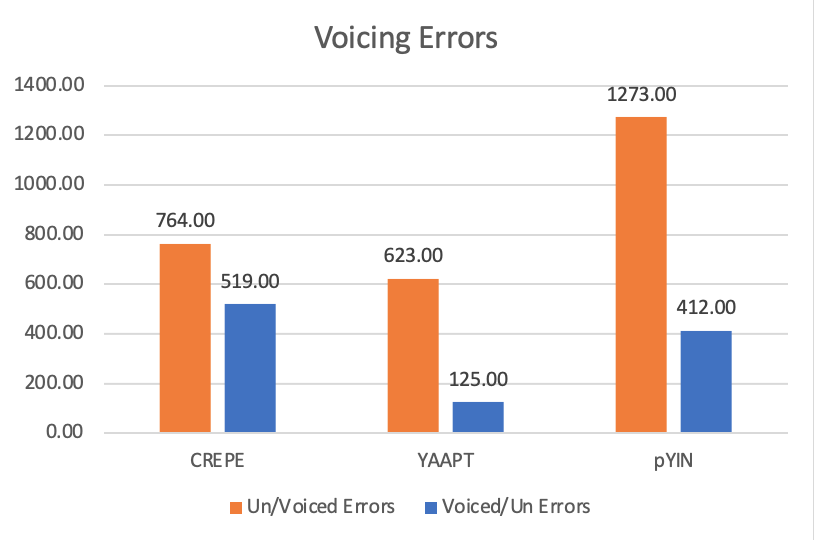}
    \caption{Voicing Errors for pYIN, YAAPT, and CREPE}
\end{figure}
\begin{figure}
    \center
    \includegraphics[width=0.45\textwidth]{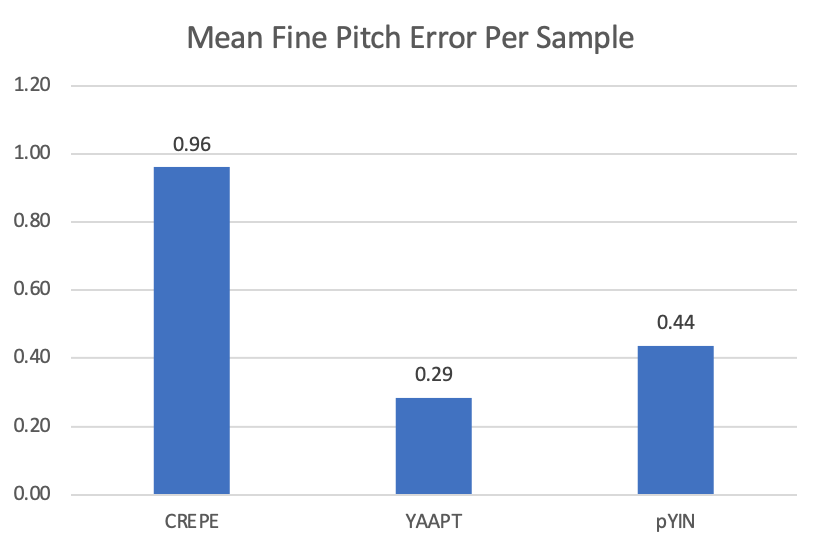}
    \caption{Mean of Fine Pitch Errors per Normalized 16k Hz Sample Resolution for pYIN, YAAPT, and CREPE}
\end{figure}
\begin{figure}
    \center
    \includegraphics[width=0.45\textwidth]{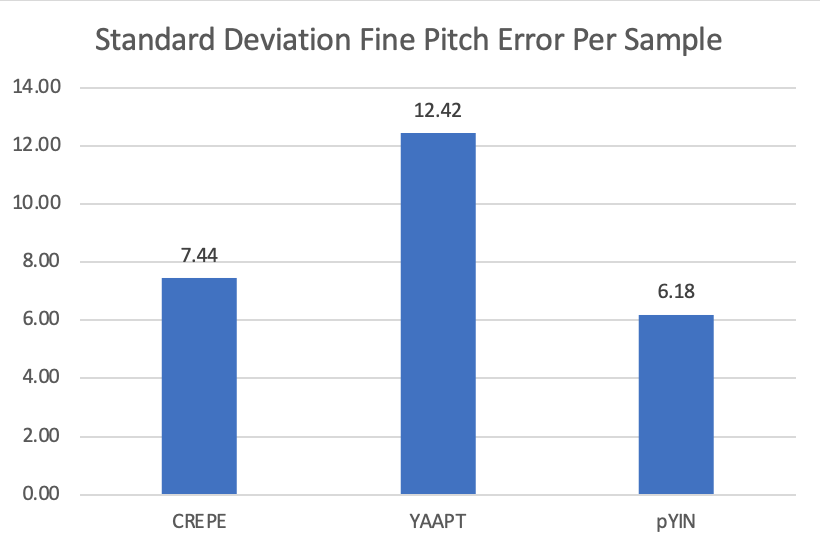}
    \caption{Standard Deviation of Fine Pitch Errors per Normalized 16k Hz Sample Resolution for pYIN, YAAPT, and CREPE}
\end{figure}

\section{Conclusion}
In this work, two conventional pitch predictor algorithms were compared with a neural net based algorithm. For the small size data base, the results show that a neural net approach can work as well as the conventional methods. Of the small number of issues with the neural net, most occurred in the voicing decisions. This can be attributed to the lack of speech signals in the test data set. Future work may include using a noisy test database to study the PDA performance in varying environments. Further work may also include comparing the algorithm complexity and delay -- important factors in real time applications.

\end{document}